\documentclass[a4paper]{article}

\usepackage{graphicx}

\begin{document}

\title{Analysis of the fine structure of HF ground backscatter and ionospheric scatter signals based on EKB radar data}

\author{I.A.Lavygin, V.P.Lebedev, K.V.Grkovich, O.I.Berngardt}

\maketitle

\begin{abstract}
The analysis was carried out of the scattered signal in the cases
of ground backscatter and ionospheric scatter. The analysis is based on
the data of the decameter coherent EKB ISTP SB RAS radar. In the paper
the signals scattered in each sounding run were analysed before their
statistical averaging. Based on the analysis and on previously studied mechanisms, 
a model is constructed for ionospheric scatter and ground backscatter signals. Within
the framework of the Bayesian approach and based on a large amount
of the data the technique and algorithm for separating these two types of signals
are constructed. The statistical analysis of the results was carried
out based on the EKB ISTP SB RAS data.
\end{abstract}

\section{Introduction}

The main approach to the study the formation, growth, and dynamics
of high-latitude ionospheric irregularities is the use of over-the-horizon
radars. The radars are usually subdivided into continuous and pulsed
radars. Continuous radars usually provide a high signal-to-noise ratio,
and a wide range of explored spatial characteristics of irregularities
due to they use a wideband sounding signal. An example of such radars
are active and passive ionosondes \cite{Ivanov_2003,Uryadov_2013}.
Pulse radars usually have a lower signal-to-noise ratio, but they
allow us to investigate not only the energy, spatial and temporal
characteristics of inhomogeneities, but their velocities and lifetimes
also. These radars include SuperDARN radars \cite{Greenwald_1995,Chisham_2007}
and radars with a similar principle of operation \cite{Berngardt_EPS}.

Due to a substantial effect of refraction on the sounding signal,
which is the basic source of the over-the-horizon operation of the
radar, the signal received by the radar consists of three parts -
noise of a different nature, a signal scattered from ionospheric irregularities
and a signal refracted in the ionosphere and scattered from the earth's
surface \cite{Milan_1997}. In addition, at large distances can exist
the signals that are consequently scattered from ionospheric irregularities
and from the earth's surface \cite{Pinnock_2002}. The physical mechanisms
responsible for scattering from the earth's surface and from ionospheric
irregularities are different, and consequently the ways of interpreting
the scattered signal are also different. However, of practical interest
is the problem of separating signals scattered from ionosphere and
scattered from the earth's surface, as well as search the characteristics
of signals that allow such a separation.

One of the main techniques used for separating groundscatter and ionospheric
scatter signals at the present time at SuperDARN radars is analysis
of their averaged spectral characteristics. It is assumed that only
groundscatter signal has sufficiently low Doppler shift and low spectral
width \cite{Blanchard_2009}. But in some cases using this technique
causes errors. From one hand, the radar measures line-of-sight velocity
and the irregularities moving across the line of sight produce zero
Doppler shift. From the other hand, when the
background ionosphere is sufficiently disturbed, the Doppler shift
can be large enough  \cite{Hayashi_2010,Grocott_2013}
and significantly higher than statistically calculated thresholds.
All these cases significantly complicates the separation problem.

Therefore various, more sophisticated methods are being developed
now to solve the problem of separation of IS and GS signals - from
a significant increase of spectral resolution by using longer sounding
sequences \cite{Berngardt_2015c}, complex spectral processing techniques 
\cite{Barthes_1998} and raytracing of radiosignal propagating
in the ionosphere \cite{Liu_2012} to a complex spatio-temporal
analysis of the areas in which the scattered signal is observed \cite{Ribeiro_2011}.
However, the physical models of GS and IS scattering in the problem
of analysing and separating these signals at SuperDARN radars apparently
were not taken into account.

The analysis of the GS and IS scattered signal is based on the data
of the pulse decameter coherent radar EKB ISTP SB RAS. The analysis
of the signals scattered in each single sounding run was carried out
without statistical averaging of power characteristics, with taking
into account their phase structure. Signal model for single sounding
run was constructed, allowing separation of GS and IS signals without
assumptions about their Doppler shift, spectral width, or additional
qualitative considerations.

\section{EKB radar observations}

Ekaterinburg coherent decameter radar (EKB ISTP SB RAS) is a CUTLASS
type radar developed at the University of Leicester\cite{Lester_2004}
and assembled jointly with IGP UrB RAS under the financial support
of the Siberian Branch of the Russian Academy of Sciences and Roshydrometeorological
Service of the Russian Federation at Arti observatory of IGP UrB RAS.
The radar transmitting and receiving antenna system is a linear phased
array. It provides a beamwidth of the $3^{o}-6^{o}$ depending on
sounding frequency and 16 fixed beam positions within $52^{o}$ field
of view. The spatial and temporal resolution of the radar is 15-45
km and 2 minutes, respectively. The frequency range of the radar is
8-20 MHz, it allows the radar to operating in over-the-horizon mode.
The radar peak pulse power 10 kW allows it to operate up to 3500-4500
km radar range. Short sounding signals provide a low (about 600 Watt)
average radar power, which allows it to operate in a 24/7 monitoring
mode.

\begin{figure}
\includegraphics[scale=0.12]{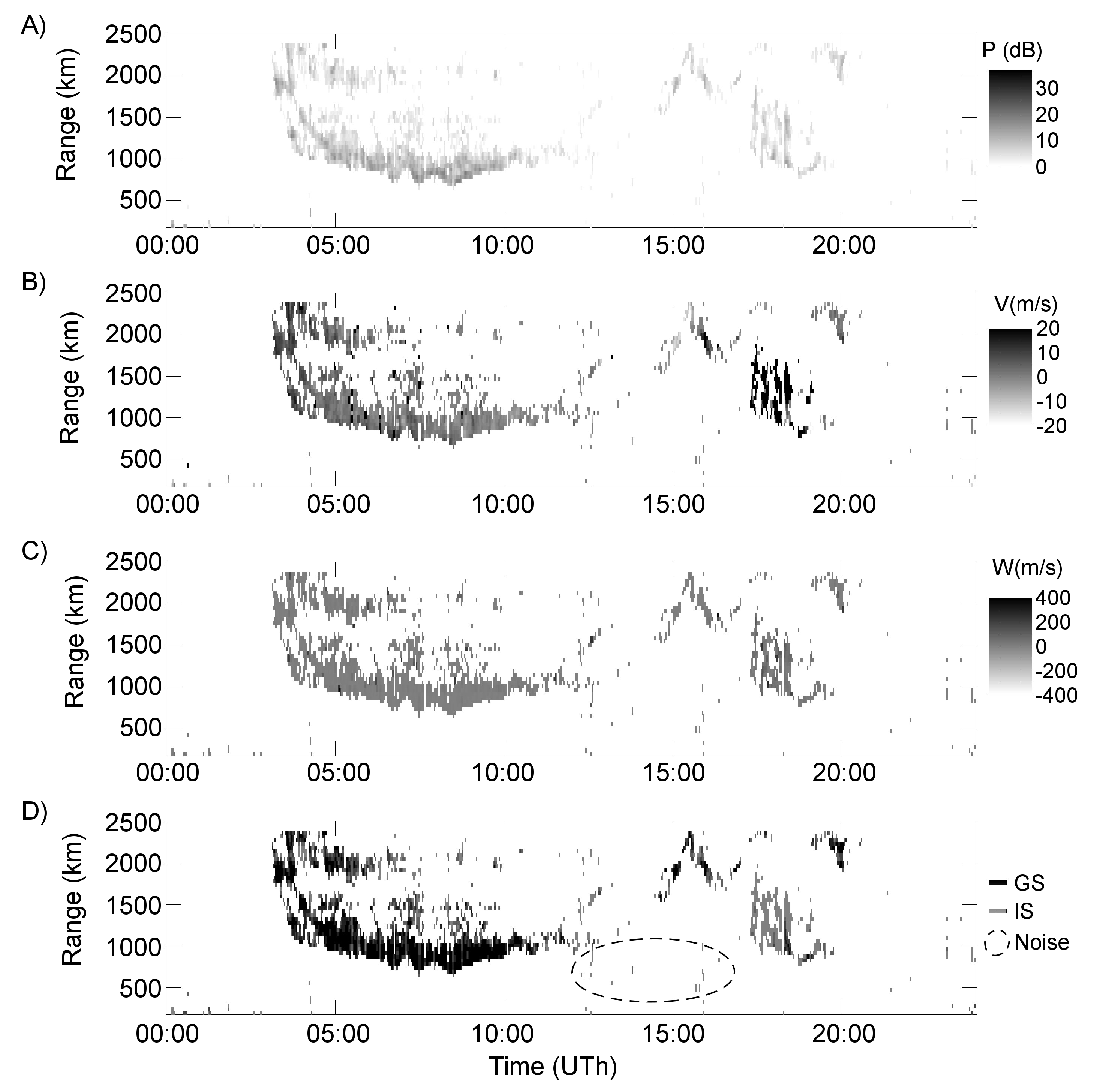}
\caption{An example of scattered signals received at EKB ISTP SB RAS radar
(at single beam). From the top to the bottom: A) is the power of the
scattered signal; B) is the Doppler velocity in the range -40 - 40
m/s; C) is the spectral width in equivalent velocity units; D) signal
types detected by standard SuperDARN program FITACF (groundscatter
flag) and noise}
\label{fig:FIG1}
\end{figure}

The standard mode of the radar operation is the measurement of the
average correlation function of the signal and using it to estimate
the scattering irregularities parameters. The basic technique of parameter
estimation is the standard FITACF program, developed and improved
by the SuperDARN community \cite{Ribeiro_2013}. The main irregularities
parameters produced by the FITACF program are the scattered signal
power, Doppler shift and spectral width, estimated in two models of
the correlation function - exponential and Gaussian \cite{Hanuise_1993}.
Fig.\ref{fig:FIG1}shows an example of the data obtained at EKB ISTP
SB RAS radar (at one of its beams). In Fig.\ref{fig:FIG1} shown the
areas corresponding to the basic kinds of scattered signals analysed
by the radar: the signal scattered from the earth's surface (groundscatter,GS),
the signal scattered from ionospheric irregularities (ionospheric
scatter, IS), scattering from meteor trails (meteor echo) and noise.
In Fig.\ref{fig:FIG1} one can see the well known basic characteristics
of the received signals: low velocities and spectral widths of GS
signals, high velocities and spectral widths of IS signals, and spatio-temporal
fragmentation and short radar ranges of the meteor echo.

The standard approach to separating signals of different types is
the separation based on the spectral parameters of the mean autocorrelation
function - the spectral width and Doppler frequency shift. In this
approach the GS signals are usually detected by small values of both
parameters, not exceeding 30-40 m/s \cite{Blanchard_2009}, and the
IS signals are detected by large values of these parameters. The problem
of separating these two signal kinds is extremely important in the
case when the ionospheric irregularities have a narrow spectrum and
a small Doppler shift, which is sometimes observed when the drift
of the irregularities is perpendicular to the radar line-of-sight.
In this case, the described technique can lead to significant failures,
and one should use more complex techniques, for example, cluster analysis
\cite{Ribeiro_2011}. Often these techniques are poorly justified
from the physical point of view. The main task of the paper is to
build a physically clear model of the scattered signal for both IS
and GS, that takes into account their phase structure and allowing
their separation.

Currently, at SuperDARN radars (similar to EKB ISTP SB RAS radar),
there is a great emphasis on measurements of the full waveform of
the scattered signal. The use of a full waveform is useful in the
studies of meteor echo \cite{Yukimatu_2002}, the digital formation
of antenna pattern \cite{Parris_2008} and in many other problems.

An essential problem useful for producing techniques for processing
scattered signals the models of the signals. A relatively standard
approach to simulating a scattered signal is the model of a large
number of random scatterers \cite{Rytov_1988,Ishimaru_1999}. As it
was shown earlier, the model corresponds well to the experimental
data \cite{Farley_1969,Moorcroft_1987,Andre_1999,Moorcroft_2004}
and can be used to producing realistic simulators of the received
data \cite{Ribeiro_2013b}, and to develop signal processing techniques
for signals accumulated over small number of sounding runs (realizations)
\cite{Reimer_2016}.

Earlier we analysed individual realizations of the signal scattered
by field-aligned ionospheric irregularities based on Irkutsk incoherent
scatter radar data. We demonstrated \cite{Grkovich_2011} that the
signal scattered by such irregularities in the VHF frequency band
can be represented as a superposition of a small number of elementary
responses. The shape of the responses repeats, in the first approximation,
the shape of the sounding signal, but has random Doppler shift and
random initial phase. The Doppler shift range is determined by the
average spectrum of the scattered signal. The checking of the adequacy
of the model for short radio waves requires a high sampling rate,
of the order of a few points per duration of the sounding signal,
and it is a critical requirement.

Currently, only a small number of SuperDARN radars have the capability
of digitization of the signal with high sampling rate, for example
\cite{Parris_2008}. Initially, the EKB
radar did not have this capability, digitizing a signal with low duty
cycle (one point per sounding pulse duration). To investigate the
fine structure of the scattered signals the radar was reprogrammed
by us to work in the mode of increased sampling frequency. The maximal
sampling frequency achieved by us in a regular mode is 5 points per
pulse duration. In a special mode with non-standard sounding sequences,
we have achieved the sampling frequency of 15 points per pulse duration
($T_{d}=20mks$ and $T_{p}=300mks$, respectively). In terms of range
resolution these values correspond to $L_{d}=3km$ and $L_{p}=45km$.This
sampling frequency does not allow us to measure the correlation function
of the signal, so it is not used in regular measurements. The use
of such a high sampling frequency makes it possible to study in detail
the phase structure of scattered signals and makes it possible to
verify the models of the scattered signal in regular mode (with 5
points per pulse duration) and special mode (with 15 points per pulse
duration).

As a criterion that determines the quality of the scattered signal
model, we have chosen its adequacy for the solution of an actual and
widely studied problem - the problem of separating groundscatter signals
(GS) and ionospheric scatter (IS).

Making special experiments significantly increases the amount of received
information, so we conducted several experiments with a temporal resolution
15 points per pulse in various geophysical conditions during October
22, November 2-4, December 8-11, 2016, January 5 and April 26-27,
2017 . The regular observations with a temporal resolution 5 points
per pulse are conducted from February 2017 till now. Further in the
paper the data obtained during these experiments are used.

The study was conducted as follows. From the obtained set of experimental
data, we manually selected two test data sets (for GS and IS signals)
to in each set there was only one intense response over range - either
first-hop groundscatter (GS) or ionospheric scatter (IS) , and we
have no significant doubt about the nature of scattering in each particular.
For GS signal we selected the region with a specific horseshoe-shaped
spatio-temporal dependence of range vs. time; for the IS, mainly we
used mostly evening and night responses and some observations of daytime
ionospheric scatter clearly separated from the groundscatter.

The search for the region of intense scattering was carried out in
the range of distances of 180-2000 km. After it is found, a range
of distances (range window) was investigated with a length $R_{w}=1000km$
(which corresponds to $T_{w}=6660\mu s$) centred at the position
of the detected intensity maximum.

\begin{figure}
\includegraphics[scale=0.5]{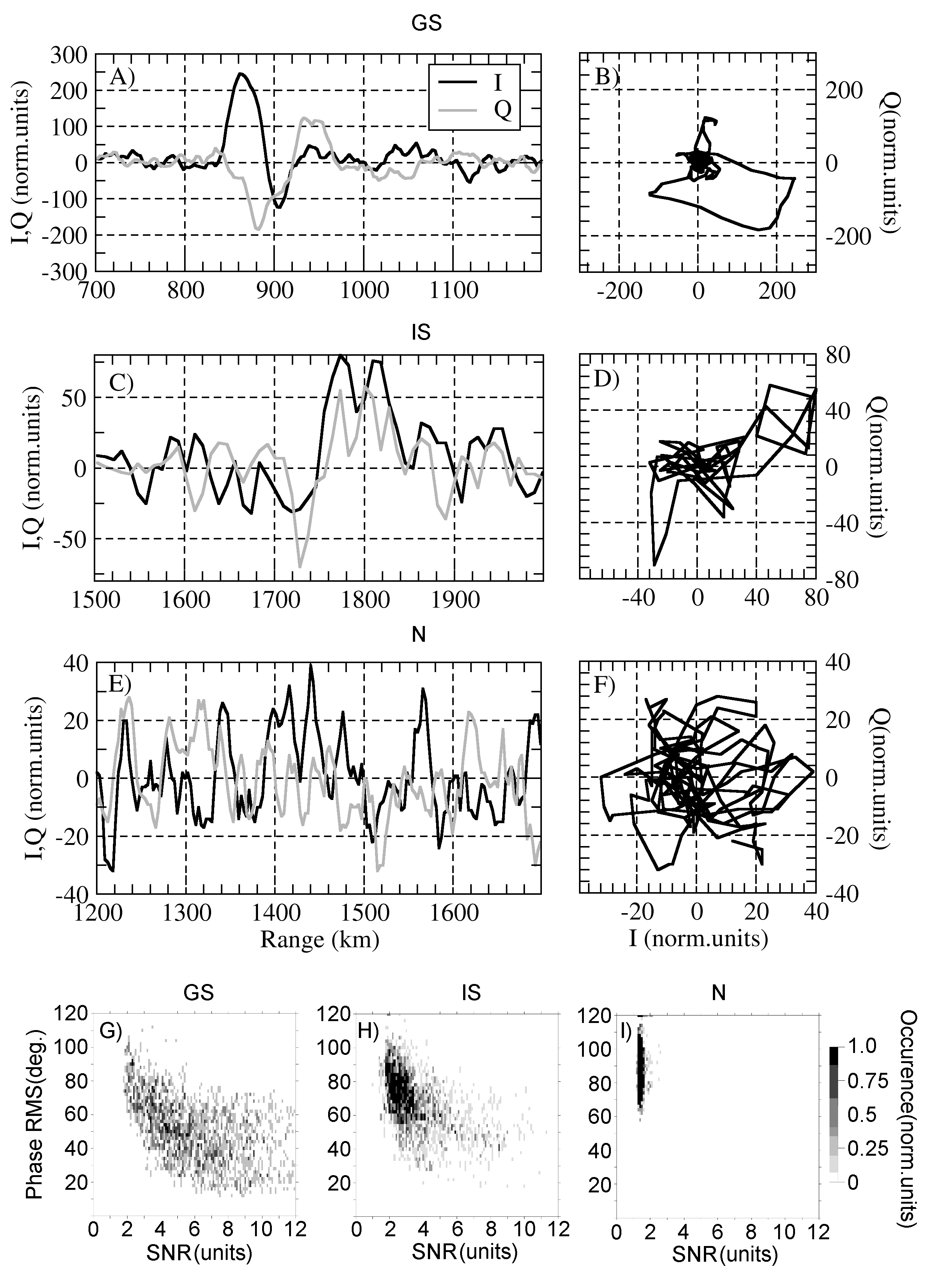}
\caption{A-F) example of realizations of the received signal (quadrature components
and phase of the signal) with a high sampling frequency in the cases
of the groundscatter (A-B), the ionospheric scatter (C-D) and the
noise (E-F). G-I) - Distributions of detected signals as a function
of signal-to-noise ratio and the phase RMS (\ref{eq:sqo_phase})
for different types of scattered signals: for groundbackscatter (G),
for ionospheric scatter(H), and for noise(I).}
\label{fig:FIG2}
\end{figure}

Fig.\ref{fig:FIG2}A-F shows examples of the received signal realizations
(their amplitude and phase structure) in the cases of the groundscatter
(Fig.\ref{fig:FIG2}A-B), the ionospheric scatter (Fig.\ref{fig:FIG2}C-D)
and the noise (Fig.\ref{fig:FIG2}E-F). The ionospheric scatter and
groundscatter is chosen with a large signal-to-noise ratio, so that
it is possible to confidently illustrate their amplitude-phase structure.
It can be seen from the Fig.\ref{fig:FIG2}A-F that both types of
scattered signals (GS and IS) have a certain phase structure, which
allows to considered them as non-random functions, in contrast to
noise (N) with nearly no phase structure. To validate this assumption,
we carried out a detailed analysis of the scattered signals on a large
amount of data.

\section{Coherent signal shapes}

Due to presence of a specific phase-amplitude structure in the received
GS and IS signals, it is of interest to parametrize this structure,
create a semi-empirical model for it, and to determine the parameters
of this model.

The GS signal has been studied for a long time in various experiments.
It is known that this signal is formed by a strong refraction of the
radiowaves in the ionosphere, leading to the focusing of radiowave
at the boundary of the dead (skip) zone (a zone at which the receiving of
radiowaves at a given frequency is impossible) \cite{Budden_1985}.
Scattering of this high-amplitude signal by the ground surface irregularities
causes a strong received signal at a range corresponding to the range
to the boundary of the dead zone. The irregularities of the Earth's
surface with scales of the order of the wavelength (tens of meters)
are nearly quasistationary (with the exception of the sea surface).
The background ionosphere at the scales of the Fresnel zone radius
(of the order several kilometres) is responsible for focusing the
radio signal and also varies relatively slowly. Therefore, the GS
signal is nearly stationary one in the first approximation, the range
to it is also nearly constant. Thus, it can be qualitatively described
from realization to realization, as a signal with an static phase-amplitude
shape at a constant range, only its initial phase can vary from realization
to realization. There are analytical expressions describing the dependence
of the power of this signal on the range \cite{Tinin_1983}. They
predict an asymmetry of its shape relative to the position of its
maximal energy, and this corresponds well with experimental observations 
at radars \cite{Bliokh_1987}. 

The amplitude-phase structure of the ionospheric scatter signal without
averaging is practically not investigated, and the conventional models
are practically absent. Only statistical models exist. In the framework
of the ionospheric scatter model considered for analogous irregularities
in the VHF band \cite{Grkovich_2011}, it was shown that the scattered
signal is in most cases can be interpreted as a single response of
the shape that repeats the sounding signal and differs from realization
to realization only by the initial phase and Doppler shift. By analogy,
let's consider the same model in HF. In addition, let's assume that
the position of the scattering region is stationary and Doppler shift
is nealy zero. The assumption of nearly zero Doppler shift does not
contradict the \cite{Grkovich_2011} model, since the sounding pulse
is short and phase changes within single pulse duration are very small
too (the possible irregularities Doppler drift velocities are within
1 km/s, for sounding frequency 10-11MHz and for sounding pulse duration
$300\mu s$ this lead to a phase variations of not more than 8 degrees).

The assumption we made about the stationarity of the scatterer position
is more strict. From qualitative considerations, it can be justified
as follows: within the radar equation in presence of refraction \cite{Berngardt_2016},
the locations of effective scattering that are equivalent to the scattering
at a point object are the locations with a given level of refraction
in which at the same time an aspect sensitivity conditions are satisfied.
Thus, their positions in the first approximation are determined by
relatively large-scale structure of the ionosphere and the magnetosphere
and have a relatively slow dynamics. Therefore, we can assume that
this assumption is also fulfilled, later we will check this from experimental
data. Thus, even in the case of different Doppler drifts (for the
characteristic velocities \textless{}1 km/s observed in the experiment),
within the framework of the \cite{Grkovich_2011} model the scattered
signal can be interpreted as a number of pulses with similar amplitude
and phase structure located at approximately the same radar range
from realization to realization.

Thus, the unified signal model that describes in the first approximation
both GS and IS signals is a signal that has unknown amplitude-phase
shape at constant range that does not change from realization to realization
and that has an initial phase varying from realization to realization.
To determine the unknown shape of such an elementary response for
both kinds of signals (GS and IS), we used the coherent accumulation
technique. This technique is based on finding in each realization
the unknown initial phase and on coherent accumulation of the signals
over the realizations with taking into account these initial phases.

The first step of the technique is to determine the position (radar
range or radar delay $\tau_{o}$) of the most powerful scattering,
carried out by searching for the maximal signal-to-noise ratio averaged
over a given number of realizations of the scattered signal (in our
case, 20 successive sounding sequences). In this case, the average
signal level is determined as the mean square of the signal amplitude
modulus over a time window equal to the duration of the elementary
sounding pulse ($300mks$). The average noise level is calculated
as the average signal level over the entire radar range (the pulse
duration refers to the entire radar range as $1/40$). The signal-to-noise
ratio calculated in this approach is insignificantly different from
real signal-to-noise ratio at its small values. At high signal-to-noise
rations it constraints the signal-to-noise ratio at the level of about
40, which improves the stability of the technique to random noise-like
bursts.

At the second stage, the parameters $(k,\psi_{0},\psi_{1},....\psi_{n})$
of model phase $\phi_{i}^{M}\left(t,k,\psi_{i}\right)$ are determined
for the phase dependence $\phi_{i}(t)$ for each of the signals.
The model $\phi_{i}^{M}\left(t,k,\psi_{i}\right)$ is:

\begin{equation}
\phi_{i}^{M}\left(t,k,\psi_{i}\right)=k\cdot(t-\tau_{o})+\psi_{i}\label{eq:lin_phase}
\end{equation}

The calculation of the parameters is made over the region limited
by the duration of the sounding pulse with the centre at the delay
$\tau_{o}$.

As already mentioned, even the strong Doppler shifts observed in the
ionosphere, at the durations of the sounding pulse order lead only
to a slight phase changes. Therefore, the use of the linear model
(\ref{eq:lin_phase}) for the phase of elementary scattering response
is sufficient and justified. The parameter $k$ is the phase distortion
factor or Doppler shift and is assumed to be the same for all implementations.
The parameter $\psi_{i}$ is the initial phase of the scattered signal
and varies from realization to realization.

All the $N_{r}+1$ parameters of the model (\ref{eq:lin_phase}) are
determined based on the minimization condition for the root mean square
(RMS) deviation of the phase $\Omega$:

\begin{equation}
\Omega=\sqrt{\frac{1}{N_{r}T_{p}}\sum_{i=1}^{N_{r}}\int_{\tau_{0}-\frac{T_{p}}{2}}^{\tau_{0}+\frac{T_{p}}{2}}\left(\phi_{i}(t)-\phi_{i}^{M}\left(t,k,\psi_{i}\right)\right)^{2}dt}=min
\label{eq:sqo_phase}
\end{equation}

Due to the fact that the model phase (\ref{eq:lin_phase}) is linear
over all the parameters, the problem (\ref{eq:sqo_phase}) reduces
to a system of linear equations and can be solved analytically.

The root-mean-square deviation, at which the minimum of the functional
(\ref{eq:sqo_phase}) is reached, determines the root-mean-square
deviation of the phase from the linear law, and can be used to verify
the adequacy of the model (\ref{eq:lin_phase}). Fig.\ref{fig:FIG2}G-I
shows the distributions of detected signals with maximal amplitudes,
as a function of signal-to-noise ratio and the phase
RMS (\ref{eq:sqo_phase}) for different types of scattered signals:
for groundbackscatter, for ionospheric scatter, and for noise. It
can be seen from the Fig.\ref{fig:FIG2}G-I that the noise is characterized
by signal-to-noise ratios about 1.5 and by phase RMS over $90^{o}$,
which indicates its quasi-random nature and approximately constant
amplitude over the range. Between the distributions of ionospheric
scatter and ground backscatter there are no specific differences except
smaller signal-to-noise ratios of ionospheric scatter.

At the third stage, all the realizations in the studied group are
rotated by their initial phases $\psi_{i}$ computed at second stage
and added together, so the signal accumulation in the region of the
maximal signal-to-noise ratio is made with nearly the same phase.
The result of accumulation is normalized to the number of realizations
in the group:

\begin{equation}
U(t)=\frac{1}{N_{r}}\sum_{i=1}^{N_{r}}u_{i}(t)e^{-i\psi_{i}}\label{eq:model_acc}
\end{equation}

The average waveform thus obtained can be used for the subsequent
analysis of the mean structure of the scattered signal.

\begin{figure}
\includegraphics[scale=0.65]{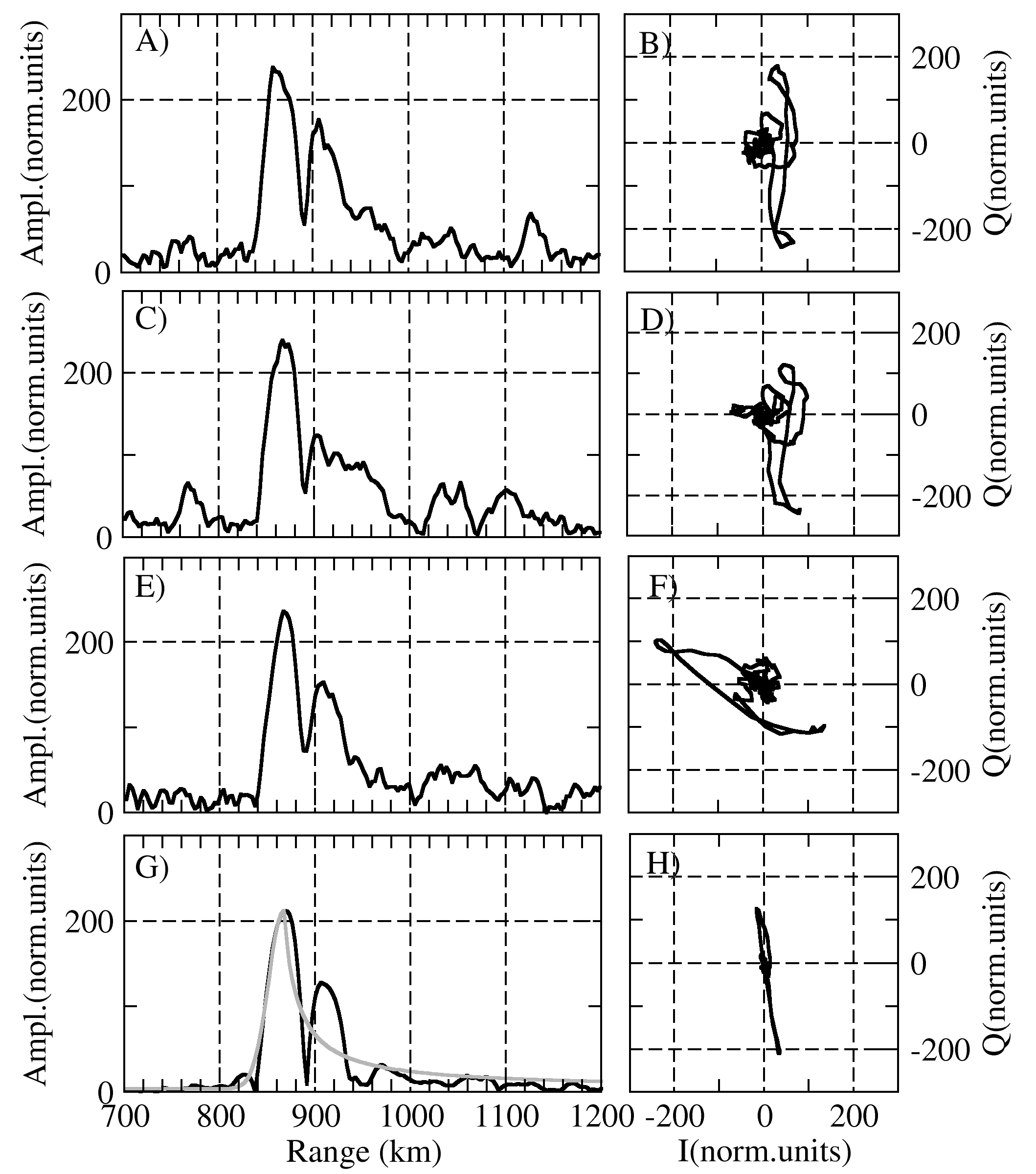}
\caption{Ground backscatter signals - amplitude (A,C,E,G) and phase diagrams
(B,D,F,H). A-F) 3 consequent realizations in the group G-H) Coherently
accumulated signal over 30 sounding pulses (\textasciitilde{}300msec).
The grey line shows the approximation of the accumulated signal by
the model (\ref{eq:model_acc}).}
\label{fig:FIG4}
\end{figure}

\begin{figure}
\includegraphics[scale=0.65]{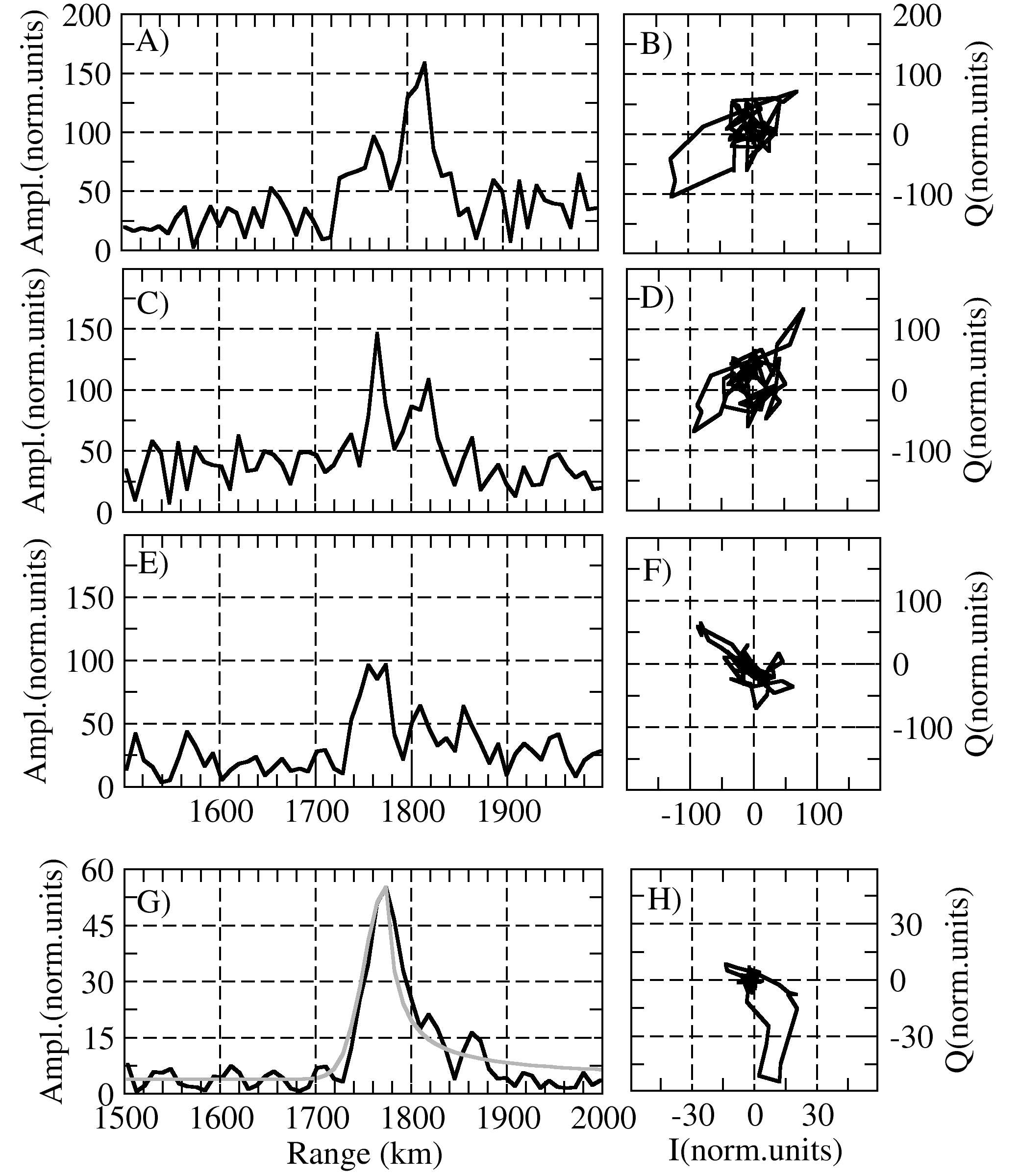}
\caption{Ionospheric scatter signals - amplitude (A,C,E,G) and phase diagrams
(B,D,F,H). A-F) 3 consequent realizations in the group G-H) Coherently
accumulated signal over 30 sounding pulses (\textasciitilde{}300 msec).
The grey line shows the approximation of the accumulated signal by
the model (\ref{eq:model_acc}).}
\label{fig:FIG5}
\end{figure}

\begin{figure}
\includegraphics[scale=0.65]{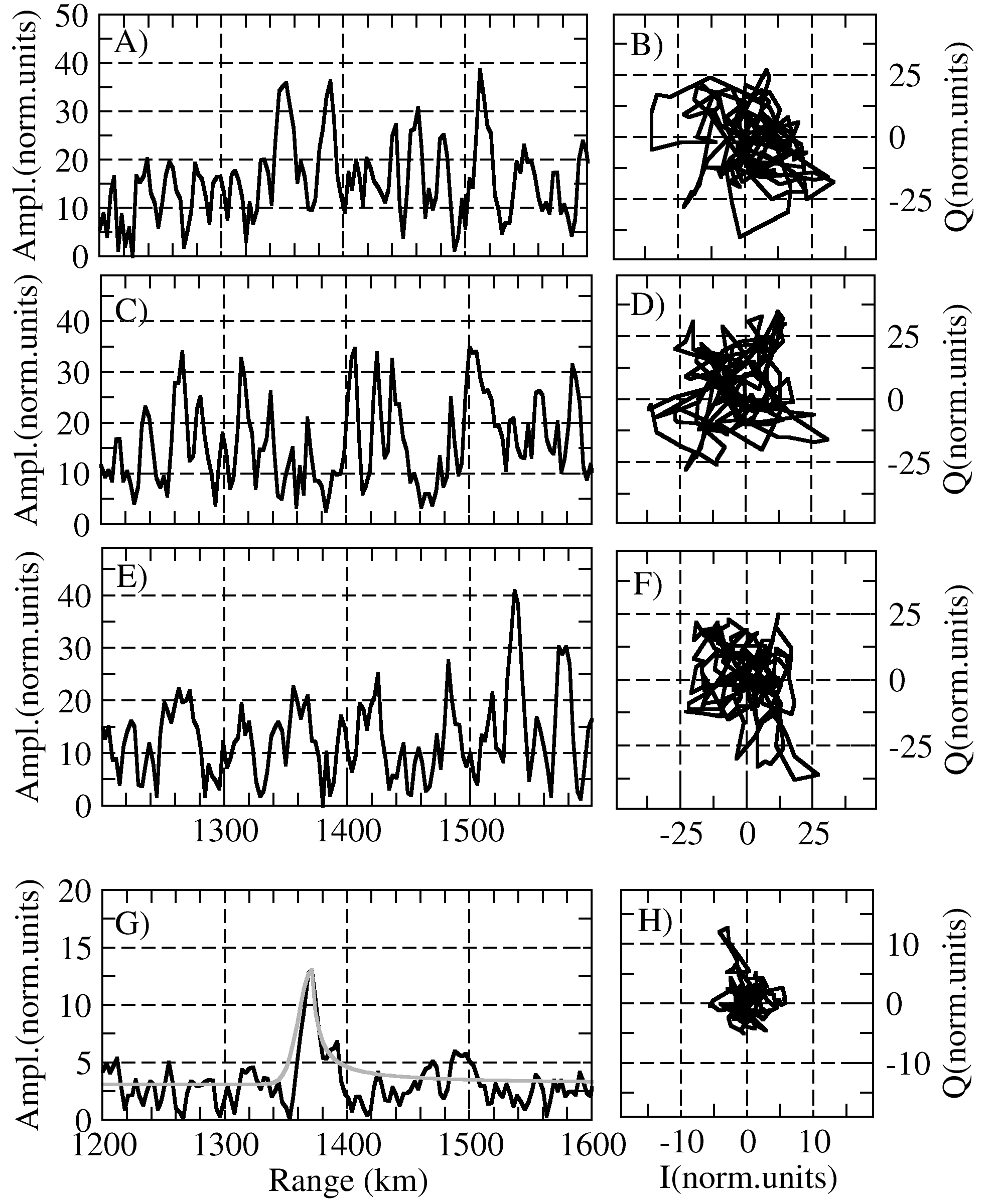}
\caption{Noise signals - amplitude (A,C,E,G) and phase diagrams (B,D,F,H).
A-F) 3 consequent realizations in the group G-H) Coherently accumulated
signal over 30 sounding pulses (\textasciitilde{}300 msec). The grey
line shows the approximation of the accumulated signal by the model
(\ref{eq:model_acc}).}
\label{fig:FIG6}
\end{figure}

In Fig.\ref{fig:FIG4}, Fig.\ref{fig:FIG5} and Fig.\ref{fig:FIG6}
shown an examples of ground backscatter signals, ionospheric scatter
signals and noise signals, as well as the mean shape of these signals.
It can be seen from the figures that the mean shape of the ground
backscatter signal, in contrast to the ionospheric scatter signal
and noise, is essentially asymmetric and has a more prolonged right
edge.

\begin{figure}
\includegraphics[scale=0.3]{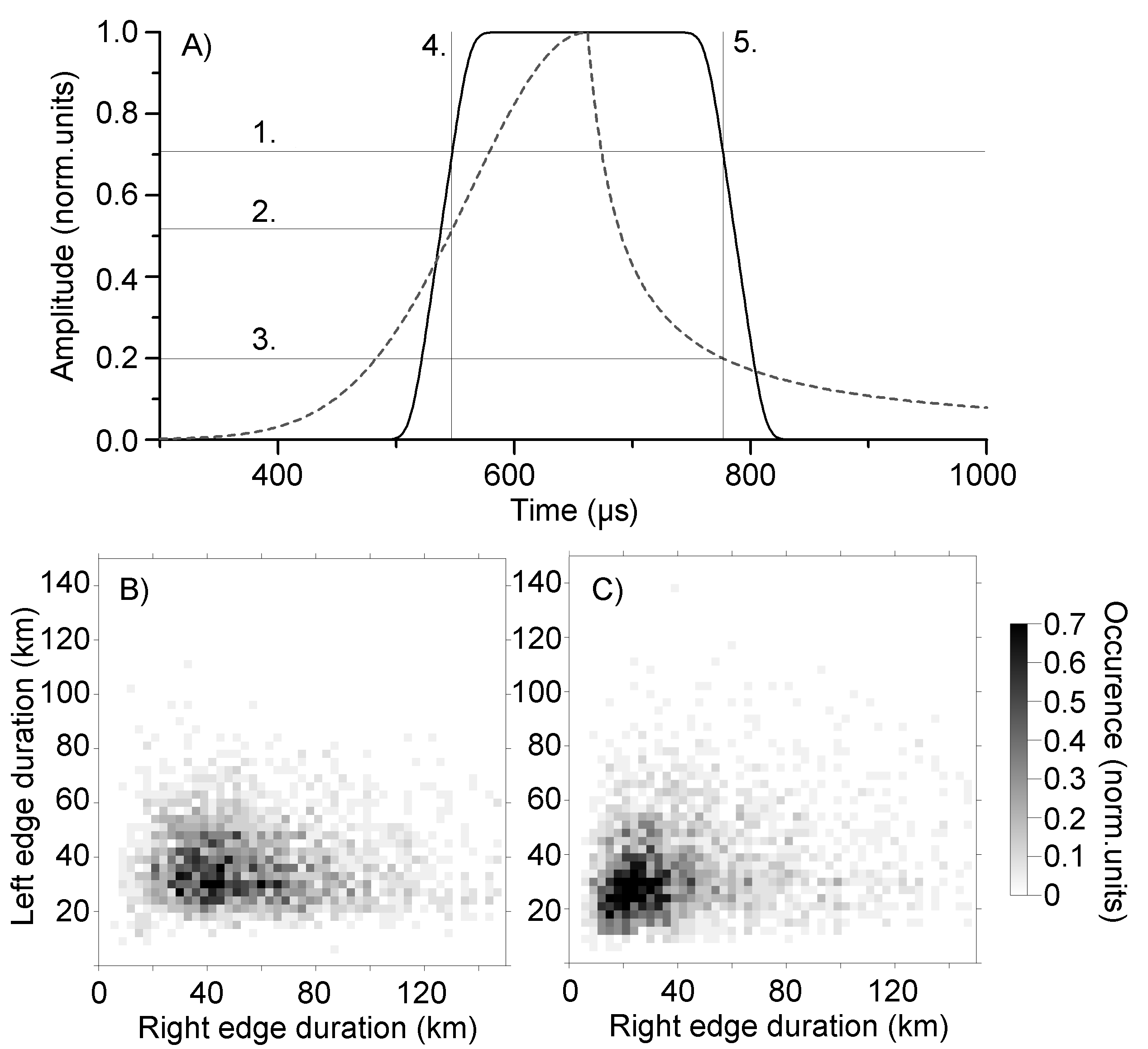}
\caption{A) Normalization of the threshold levels of the approximating functions
based on the sounding pulse. marks 1-5 are the following: 1 is level
of half power (0.7 in amplitude) of the sounding signal. 2 is the
threshold level for approximating the left edge. 3 is the threshold
level for approximating the rigth edge; 4 is the moment corresponding
to half power on the left edge of the sounding signal; 5 is the moment
corresponding to half the power on the right edge of the sounding
signal. B-C) are the distributions of edge durations of the received
signals B) is the distribution for ground backscatter; C) is the distribution
for ionospheric scatter.}
\label{fig:FIG7}
\end{figure}

To check the regularity of the observed feature, we made an algorithm
to determine of the duration of the right and left edges of the accumulated
signal and applied it to all the investigated observations. Obviously,
the first step to estimating the edge duration is to create an unified
model that can approximate both the GS signal and the IS signal. We
used the following asymmetric model (illustrated at Fig.\ref{fig:FIG7}A):

\begin{equation}
a(t,\{A,B,C\})=N+
\left\{
\begin{array}{l}
Ae^{-\left(\frac{^{t-\tau_{0}}}{C}\right)^{2}};  t<\tau_{0}\\
\frac{A}{1+(t-\tau_{0})/B};  t>\tau_{0}
\end{array}
\right.
\label{eq:a_model}
\end{equation}

where $\tau_{0}$, $A$ are the radar delay to the maximal amplitude
of the scattered signal and the maximal amplitude of the accumulated
signal correspondingly; $N$ is the noise level, defined as the sum
of the constant noise level and its RMS (determined at large distances,
where the groundscatter and ionospheric scatter signals are absent);
$B$ and $C$ are the parameters to be estimated, that characterize
the duration of the left and right edges, respectively. The choice
of this model can be qualitatively justified by the form of the asymptotic
solution for the GS power, characterized by a sharp left edge, and
by the smooth right edge \cite{Tinin_1983}. The comb structure of
the accumulated signal, as well as its phase structure has not been
investigated in the paper. In order to speed up the calculations,
the search for $B$ and $C$ values was made based on integral approach:
the integral on the left and right of $\tau_{0}$ of the experimental
signal shape should be equal to the corresponding integral of the
model function:

\begin{equation}
\left\{
\begin{array}{l}
\int_{\tau_{0}}^{\infty}(a(t,\{A,B,C\})-N)dt=\int_{\tau_{0}}^{\infty}(U(t)-N)dt \\
\int_{0}^{\tau_{0}}(a(t,\{A,B,C\})-N)dt=\int_{0}^{\tau_{0}}(U(t)-N)dt
\end{array}
\right.
\end{equation}

After determining the parameters of the model $B,C$ it is necessary
to calculate from them the durations of left and right edges correspondingly.
The calculation should be made by the special way to the model time-symmetric
sounding signal has the same right and left edges the sum of which
is equal to actual sounding signal duration. 

To migrate from the parameters of model $B$ and $C$ to the edge
durations $T_{R},T_{L}$, we determined at which threshold levels
$\varepsilon_{L,R}$ the two functions: $e^{-\left(\frac{^{t-\tau_{0}}}{C}\right)^{2}}$
and $\frac{1}{1+(t-\tau_{0})/B}$ intersect with the model shape of
a Gaussian signal signal with standard $300mks$ duration. The values
of these threshold levels are found to be $\varepsilon_{L}=0.5$ for
the left edge and $\varepsilon_{R}=0.2$ for the right edge. Thus,
the duration of the edges $T_{L},T_{R}$ are defined by us as the
distance from $\tau_{0}$ to the points at which the right or left
edge of the model function reach the right or left threshold level:
\begin{equation}
\left\{
\begin{array}{l}
a(t-T_{L},\{A,B,C\})=\varepsilon_{L}A+N \\
a(t+T_{R},\{A,B,C\})=\varepsilon_{R}A+N
\end{array}
\right.
\end{equation}

The edge durations $T_{L},T_{R}$ determined in this approach have
a clear physical sense - when the model (\ref{eq:a_model}) is fitted
into a real sounding signal $a_{0}(t)$ of $300\mu s$ duration, the
calculated edge durations $T_{L},T_{R}$ will be equal to $150mks$
each (half of the sounding pulse duration). An explanation of this
method for estimating the edge duration is illustrated in Fig.\ref{fig:FIG7}.
Therefore, the obtained values of $T_{L},T_{R}$ can be used to quantitatively
compare the durations of the right and left edges with the duration
of the sounding signal, and thus allow us to estimate duration of
the scattered signal edges directly in kilometres or microseconds.

By using the technique described above, we calculated the statistical
distributions of the durations of the left and right edges for IS
and GS signals, based on the the available experimental data (Fig.\ref{fig:FIG7}B-C).
From the figure one can see that the characteristics of the scattered
signals of different kinds are significantly different: the accumulated
ground backscatter signal (Fig.\ref{fig:FIG7}B) is more asymmetric
and has smoother right edge in comparison with the right edge of the
ionospheric scatter signal (Fig.\ref{fig:FIG7}C). At the same time,
the accumulated ionospheric scatter signal has relatively symmetrical
edges. This does not contradict the previously developed models for
GS\cite{Tinin_1983} and IS \cite{Grkovich_2011} signals, and validates
the use of these models in the problem under consideration.

~

\section{\emph{Coherent signal lifetimes}}

Traditionally, the HF signal scattered from ionospheric irregularities
is interpreted as a superposition of scattering from the large number
of elementary scatterers \cite{Farley_1969,Moorcroft_1987,Andre_1999,Moorcroft_2004}.
In the previous section, we showed that in the case of scattering
by field-aligned irregularities it can be interpreted as a result
of scattering by a small number of elementary scatterers spaced in
range, which is closely related to the model we proposed earlier in
VHF \cite{Grkovich_2011}.

Let's estimate the characteristic lifetimes of such elementary scatterers
from the scattered signal. To do this let's find the dependence of
the normalized cross-correlation coefficient between two different
realizations, as a function of the delay between them. Following to
the approach described before the calculation of the correlation coefficient
is made over a region of maximal signal-to-noise ratio, centred at
delay $\tau_{0}$. The duration of the region is determined
from the statistics of coherently accumulated signals - as the maximal
expected durations of the right and left edges $T_{R,0},T_{L,0}$.

\begin{equation}
R(i)=max_{n}\left\{ R_{n,n+i}(t)=\frac{\int_{\tau_{0}-T_{L,0}}^{\tau_{0}+T_{R,0}}u_{n}(\tau)u_{n+i}^{*}(\tau+t)d\tau}{\sqrt{\int_{\tau_{0}-T_{L,0}}^{\tau_{0}+T_{R,0}}|u_{n}(\tau)|^{2}d\tau\int_{\tau_{0}-T_{L,0}}^{\tau_{0}+T_{R,0}}|u_{n+i}(\tau)|^{2}d\tau}}\right\} \label{eq:K_corr}
\end{equation}

where $u_{n}^{*}$ is the complex conjugate value of the signal $u_{n}$
received in n-th sounding run.

As one can see in Fig.\ref{fig:FIG4},\ref{fig:FIG7}B, the duration
of ground backscatter signal in a single realization (before its coherent
accumulation) can reach up to 200-300 km. At the same time, the duration
of the left edge for both kinds of signals usually does not exceed
60 km ($400\mu s$, see Fig. \ref{fig:FIG7}B-C).

Therefore, we chose the following values for the duration of the right
and left edges used for calculation of correlation coefficient (\ref{eq:K_corr})
$T_{R,0}=400mks$, $T_{L,0}=1600mks$.

For a detailed analysis of the irregularities lifetime we developed
an algorithm for calculating the correlation coefficient at arbitrary
moments, both comparable with and exceeding the duration of the sounding
sequence (70 ms). The basis of the algorithm is the calculation of
the correlation coefficient at delays (lags) corresponding to the
combination lags between the sounding pulses. The main property of
the sounding sequences, based on the properties of Golomb rulers,
is that the combination lags between different sounding pulses are
always different and practically uniformly cover the region of lags
within the duration of the sounding sequence. The correlation coefficient
for signals at such combinational delays makes it possible to determine
the dependence of the correlation coefficient at small lags that are
shorter than the sequence length.

Analysis of the correlation coefficient at lags exceeding the duration
of the sounding sequence is traditionally not carried out at SuperDARN
radars and at EKB radar. Most often this is associated with the complexity
of the end-to-end synchronization of all sounding sequences. In the
approach we proposed, this is possible. To evaluate the correlation
coefficient at large lags exceeding the duration of the sounding sequence
we calculated it at lags equal to the delay between the response from
the first pulse of the first sounding sequence and the pulses of all
subsequent sounding sequences. This approach allows us to obtain a
detailed dependence of the correlation coefficient at nearly arbitrary
lags.

\begin{figure}
\includegraphics[scale=0.4]{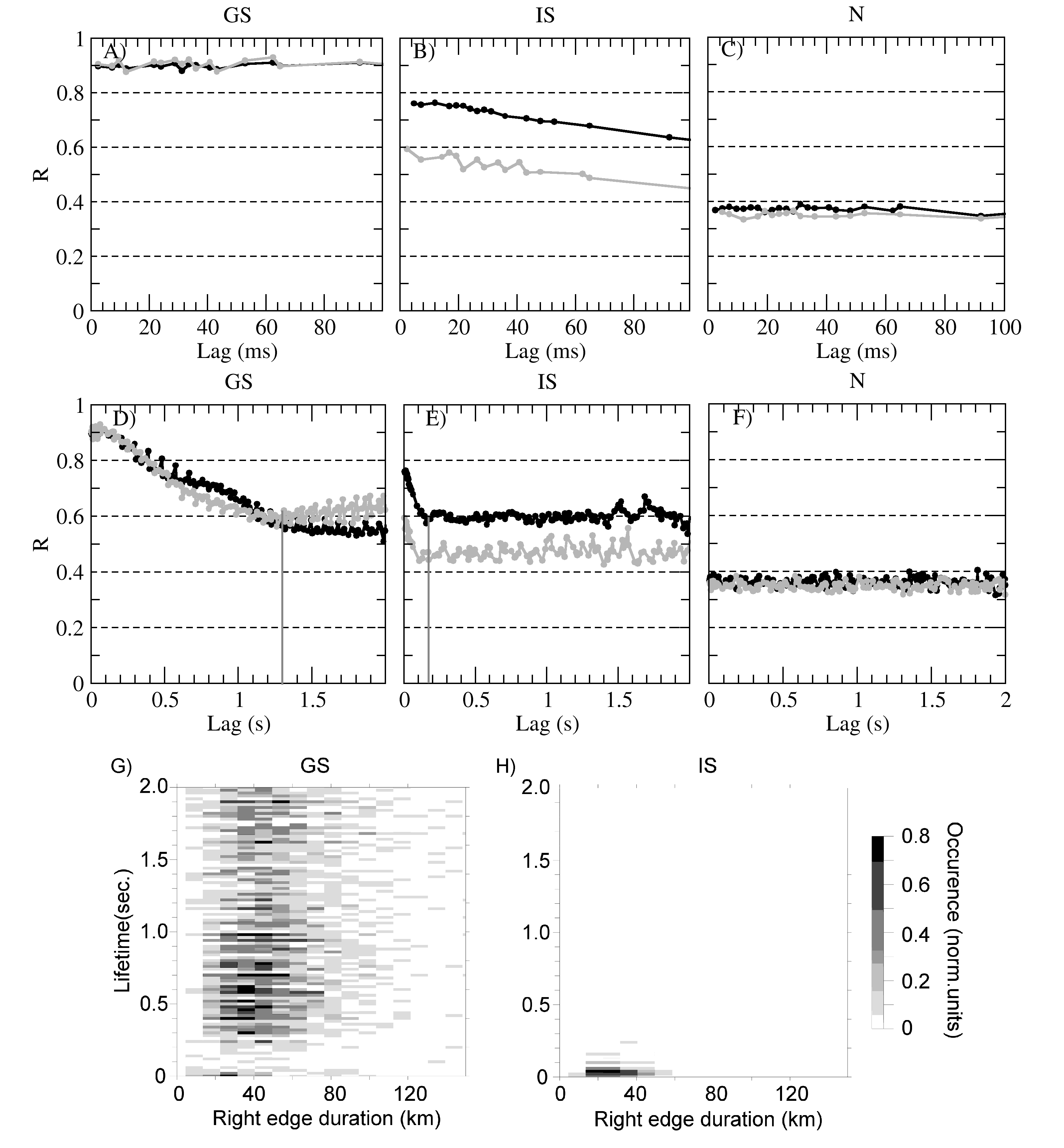}
\caption{A-F) - correlation coefficient for different signal types. A-C) are
is correlation coefficient at small lags comparable with sounding
sequence duration for groundscatter(A), for ionospheric scatter(B)
and for noise (C); D-F) is correlation coefficient at large lags comparable
with averaging interval in regular sounding mode, for groundscatter
(D), for ionospheric scatter (E) and for noise (F). Grey veritcla
line at D-E corresponds to coherent signal lifetime. G-H) - distributions
of signals over the scatterer lifetime (in seconds) and the right
edge duration (in km.): G) is the distribution for ground backscatter;
H) is the distribution for ionospheric scatter.}
\label{fig:FIG8}
\end{figure}

Examples of the algorithm functionality for different kinds of scattered
signals are shown in Fig.\ref{fig:FIG8}. It can be seen from the
fig.\ref{fig:FIG8}A-C the groundscatter and ionospheric scatter signals
differ significantly from the noise - they have higher correlation
coefficient at small lags. The correlation coefficient for IS signals
increases at small lags, and this allows to interpret the IS signal
as a result of scattering by scatterers with a relatively short lifetime
(hundreds of milliseconds). This does not contradict the existing
data on the lifetime of instabilities that form field-aligned irregularities
\cite{VILLAIN_1996}. It should be noted that when the signal noise
ratio decreases, this property still persists, although it becomes
less pronounced, and the maximum correlation coefficient at small
lags decreases. 

Ground backscatter signals also tends to decrease the correlation
coefficient with a lag, but the characteristic rate of the decrease
is much lower. From Fig.\ref{fig:FIG8}A-B one can
see, that in some cases (black lines) it is difficult to differ groundscatter
from ionospheric scatter using lifetime (small IS spectral widths
case) at lags, provided by standard sounding sequences. At other cases
(green lines in \ref{fig:FIG8}A-B) they can be differed (big IS spectral
widths case).

Analysis of correlation at extra large lags, compared with whole averaging
interval for regular sounding is shown in Fig.\ref{fig:FIG8}D-F.
It can be seen from Fig.\ref{fig:FIG8}D-F that when analysing extra
large lags, the average lifetime of GS signals (\textgreater{}1s)
exceeds the lifetime of IS signals(\textless{}250ms).

This can be explained by the physics of their formation - GS signal
is a signal scattered by nearly stationary ground surface inhomogeneities,
and its nonstationarity is mainly related with the existence of medium-
and large-scale ionospheric irregularities that affect the refraction
of this signal and, accordingly, its decorrelation with time. On the
other hand, it is known that the lifetime of individual small-scale
inhomogeneities is small, and can be estimated from the spectral width
of the IS signals (
rarely exceeds $250ms$. Thus, the lifetime data obtained by us do
not contradict the known characteristics of the scattered signals
and the physics of their formation.

To automaticaly calculate the lifetime we use the following technique.
In Fig.\ref{fig:FIG8}D-E one can see that correlation coefficient falls 
with delay ('lag') and reaches a certain stable ('noise') level.
Therefore we define the lifetime as delay at which the correlation 
coefficient is bigger than certain threshold level $R_{th}$. 
This level is calculated over large lags, as $R_{th}=<R>+\Delta R$. Here

\begin{equation}
\begin{array}{l}
<R>=\frac{1}{T_2-T_1} \int_{T_1 = 2sec}^{T_2 > T_1} R(t) dt \\
\Delta R= \sqrt{ \frac{1}{T_{2}-T_{1}} \int_{T_1 = 2sec}^{T_2 > T_1} (R(t)-<R>)^2 dt}
\end{array}
\end{equation}

\section{Separation of IS and GS signals}

As was shown earlier in the paper, GS and IS signals have different
characteristics of the mean shape of the scattered signal and different
lifetime (correlation) of the elementary scatterer. This allows us
to construct effective techniques for separating these signals by
their amplitude-phase and correlation characteristics.

One of the standard approaches to signal separation is the method
for testing statistical hypotheses \cite{Lehman_2005}. This method
reduces the problem of separating signals to the problem of determining
the detection boundary shape in the multidimensional space of signal
characteristics, on the one side of which the signal will be considered
as GS signal, and on the other side of the boundary it will be considered
as IS signal. There are several methods of making such a boundary
shape, and they are based on minimizing the sum of the errors of the
first and second kind (errors of incorrect acceptance and incorrect
rejection of the hypothesis). We used the simplest Bayesian inference,
under assumption of the equiprobability of IS and GS signals.

As it was shown earlier in the paper, the basic characteristic parameters
that allow separation are the scatterer lifetime (correlation time)
and the duration of the right edge of the coherently accumulated signal.
Fig.\ref{fig:FIG8}G,H shows the distributions of these GS and IS
signal characteristics - the right edge duration (in km.) and the
scatterer lifetime (in seconds). It can be seen from the figure that
in these coordinates, the IS signals area corresponds to a small region
concentrated near small lifetimes (y-axis) and edge durations(x-axis)
oriented along the x-axis. The most part of the GS signals are outside
this region. So the Bayesian approach should provide effective solution
for separation.

To separate these kinds of signals, we used the distribution of their
characteristics in a three-dimensional parameter space (the duration
of the right edge, the duration of the left edge, and the scatterer
lifetime). Earlier it was shown (fig.\ref{fig:FIG8}G-H) that the
IS signal distribution lies within a closed region in the parameter
space bounded by a certain surface around the coordinates centre (small
lifetimes, short right edges). The GS signal distribution in this
parameter space lies outside this surface (large lifetimes, long right
edges). In the first approximation, there is no significant correlation
between the duration of the right edge and the scatterer lifetime
(Fig.\ref{fig:FIG8}G-H), as well as between the left and right edges
of the coherently accumulated signal (Fig.\ref{fig:FIG7}). Therefore,
in the first approximation, we can use separation boundary as the
surface of a certain ellipsoid with the axes along the coordinate
axes:

\begin{equation}
\frac{x^{2}}{a^{2}}+\frac{y^{2}}{b^{2}}+\frac{z^{2}}{c^{2}}=1
\label{eq:elips}
\end{equation}

The coordinates x, y and z in our case are the scatterer lifetime,
the duration of the left edge and the duration of the right edge of
the coherently accumulated signal, respectively. The elipsoid axis
sizes - $a,b,c$ are to be determined.

To determine these parameters, three-dimensional discrete distributions
$P_{IS}(x,y,z)$ and $P_{GS}(x,y,z)$ were constructed from two experimental
data sets (for GS and for IS). The step of the discretization of distributions
over the scatterer lifetime was 2.5~ms, and over
the left and right edges - 3~km. The search for
the optimal values of the parameters $a,b,c$ was made numerically,
by a direct search over the grid by the parameter (in steps of 2.5~ms 
for $a$, and 3~km for $b$ and $c$). The condition
for optimality of the parameter set was the Bayesian criterion in the
form of minimization of the functional of the total error of the first
and second kinds:

\begin{equation}
\Omega=\int_{r\left(\theta,\alpha\right)>\rho_{d}\left(\theta,\alpha\right)}P_{IS}(x,y,z)dxdydz+\int_{r\left(\theta,\alpha\right)<\rho_{d}\left(\theta,\alpha\right)}P_{GS}(x,y,z)dxdydz=min
\label{eq:elips_omega}
\end{equation}

Integration in the Cartesian coordinate system is made over the region
outside the surface of the ellipsoid in the first term, and over the
region inside this ellipsoid - in the second term. Physically, this
condition corresponds to the fact that most of the values of the scattered
signal characteristics for ionospheric scatter lie inside the surface
of the desired ellipsoid, and most of the values of the scattering
characteristics of ground backscatter are outside the surface of this
ellipsoid.

The coordinates of the ellipsoid in the calculations are given in
the polar coordinate system $\left(r,\theta,\alpha\right)$. This
makes it convenient to determine the condition if test point lies
inside or outside the surface of the ellipsoid. In this case, it reduces
the separation problem to checking the conditions $r\left(\theta,\alpha\right)>\rho_{d}\left(\theta,\alpha\right)$
and $r\left(\theta,\alpha\right)<\rho_{d}\left(\theta,\alpha\right)$,
where $\rho_{d}\left(\theta,\alpha\right)$ is the equation of the
ellipsoid surface in polar coordinates.

\begin{figure}
\includegraphics[scale=0.12]{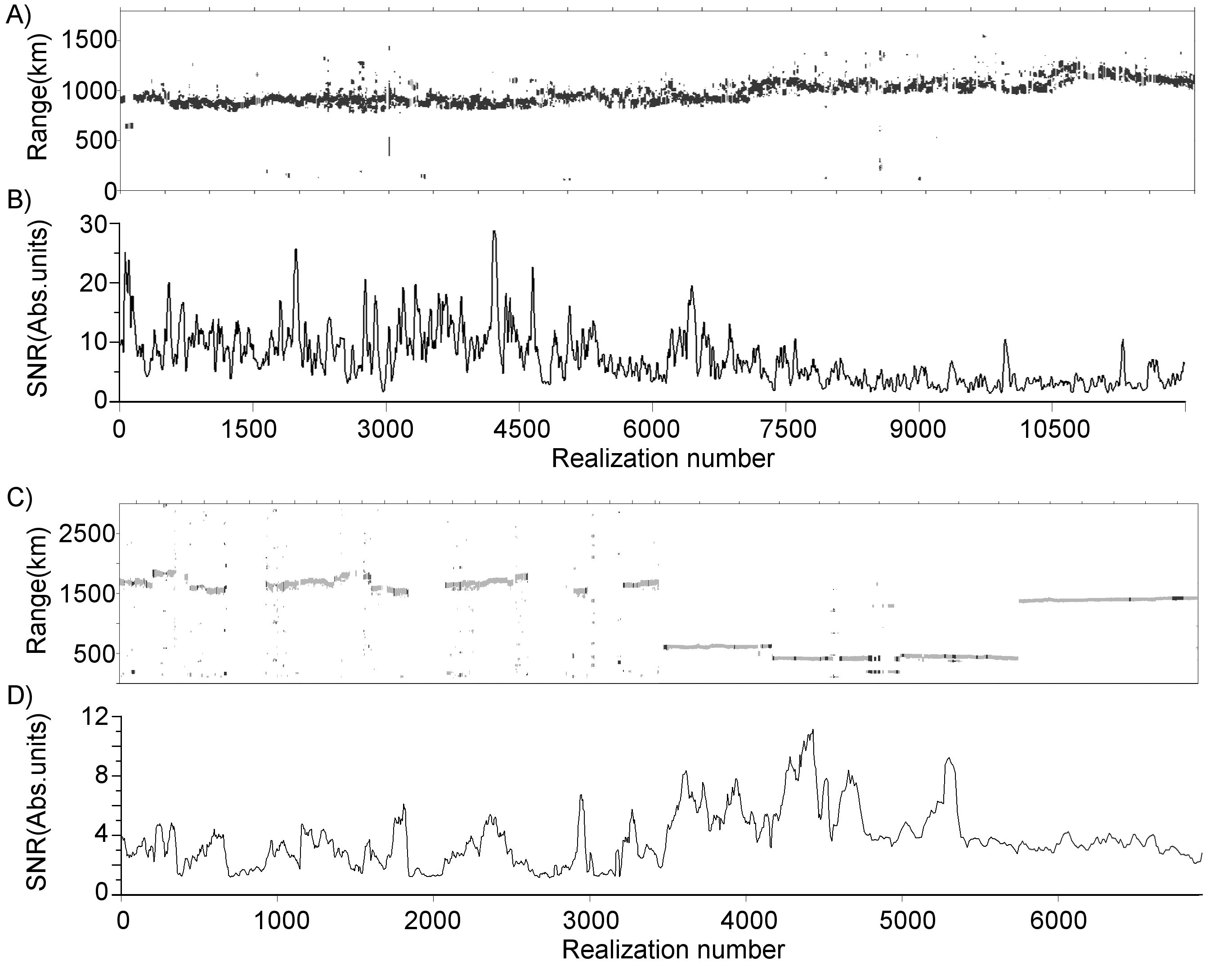}
\caption{
The results of detecting signal type by the new detection algorithm. 
Black color at (A,C) marks the signals, identified as groundscatter, grey color at (A,C)
marks the signals identified as ionospheric scatter. A) - results of processing 
GS data; B) - peak signal-to-noise ratio of GS data; C) - results of 
processing IS data; D) is the peak signal-to-noise ratio of IS data.
}
\label{fig:FIG10}
\end{figure}

Based on the learn data set with more than 13 thousand realizations and optimum condition (\ref{eq:elips_omega}) 
we calculated the following separation boundary (\ref{eq:elips}) parameters: 
a = 285 ms, b = 120 km, c = 429 km.
We use about 19 thousand realizations to test the technique (13 thousand from learn data set and 6 thousand of other 
realizations) to verify the technique. The results are shown in Fig.\ref{fig:FIG10}. As shown our analysis, the 
accuracy of groundscatter detection is about 95.1\%. Accuracy of ionospheric scatter is about 88.6\%. The total 
detection error is about 16\%. 

To illustrate the technique in Fig.\ref{fig:FIG10} we shown an example of detection of  
the type of signals for GS and IS signals based on the test data set described above. 
As one can see in Fig.\ref{fig:FIG10} in most cases the algorithm works correctly 
and the correctness of the response does not depend on the signal-to-noise ratio, 
which is an indirect sign of the validity of the developed model and the separation 
technique.


\section{Conclusion}

The analysis is made of the fine structure of decameter signals scattered
by irregularities of the earth's surface and field-aligned ionospheric
irregularities. To carry out such an analysis with a high sampling
frequency, the software of the EKB ISTP SB RAS radar was substantially
modernized. A large number of experimental data with an increased
sampling frequency has been obtained. As a result of the experimental
data analysis it is shown that signals scattered by both mechanisms
have a specific phase structure and a nonzero lifetime. This allows
us to interpret them as signals scattered by a small number of localized
irregularities with a finite lifetime. A method was implemented for
detecting the shape of the elementary response. Empirical model is
developed that allows to describe both types of signals and to determine
their characteristics - the lifetime, the duration of the right and
left edges. Statistical features of both types of scattered signals
are estimated and presented. Differences in the shape of signals scattered
from the earth's surface and from the ionosphere are detected: the
different durations of the right front of the signal and different
lifetimes. Based on the analysis of the full waveform and the Bayesian
inference approach, a method for the optimal separation of ground
scatter and ionospheric scatter signals was constructed. The technique
works without statistical averaging and without using traditional
SuperDARN methods for estimating scattered signal parameters (FITACF).
The effectiveness of the method is estimated based on EKB radar data, 
it shows total detection error about 16\%.

\section{Acknowledgements}

The work was supported by the RFBR grant \# 16-05-01006a. The functioning of
the EKB ISTP SB RAS is supported by the FSR program II.12.2.3. The data
of EKB radar is a property of ISTP SB RAS, contact berng@iszf.irk.ru

\section{References}
\bibliographystyle{plain}

\end{document}